\documentclass[12pt]{article}
\usepackage{graphicx}
\usepackage{amsmath,amssymb}
\makeatletter
\def\@cite#1{\textsuperscript{#1}}
\def\@biblabel#1{#1.}
\makeatother

\title{\bf Iterative conformal mapping approach to diffusion-limited 
aggregation with surface tension effect}

\author{Hiroshi Miki and Haruo Honjo\\
Department of Applied Science for Electronics and materials,\\
Interdisciplinary Graduate School of Engineering Sciences,\\
Kyushu University, Kasuga, Fukuoka 816-8580, Japan}

\date{}%
\begin{document}
\maketitle
\begin{abstract}
We present a simple method for incorporating the  surface tension effect 
into an 
iterative conformal mapping model of two-dimensional diffusion-limited 
aggregation. A curvature-dependent growth probability is introduced and 
the curvature is given by utilizing the branch points of a conformal 
map. The resulting cluster exhibits a crossover from compact to 
fractal growth. In the fractal growth regime, it is confirmed, by the 
conformal map technique, that the fractal dimension of its area and perimeter 
length coincide. 
\end{abstract}

{\it Keywords}: Pattern formation, Diffusion-limited growth, Conformal mapping,
Surface tension effect
\newpage

\section{INTRODUCTION}
Since its introduction by Witten and Sander in 1981\cite{WS}, 
the diffusion-limited aggregation (DLA) has been one of the most 
important models of growth processes which are dominated by 
diffusion process or, in a wider sense, of non-equilibrium and non-linear 
statistical physics and stochastic processes. The dynamical rule of 
this model is very 
simple: Initially a seed particle is placed at the center. A random 
walker is released far from the cluster and when once it touches the 
cluster, it stops walking and becomes a part of the cluster. Then next 
walker is released, $\cdots$. The process repeats again and again.

The cluster grown under the above-described process forms a fractal 
structure having several, typically five for a two-dimensional cluster, 
main branches and each branch has many 
subbranches hierarchically. Growth patterns like this have been 
observed for many kinds of experiments: electrodeposition\cite{Mat}, 
viscous fingering\cite{HS,ST,NDS}, crystal growth with its anisotropy  
suppressed\cite{HOM}, bacteria colony growth\cite{FM}, and so forth. 
These experiments, computer simulations, and theoretical studies 
have clearified the 
properties of the DLA, but some fundamental problems to be 
understood remain. 

About fifteen years ago, a novel approach was proposed\cite{HL} for 
two-dimensional DLA. 
This approach utilizes iterative conformal map, 
which automatically satisfies the Laplace equation, 
since the DLA is dominated by the field which satisfies it. 
This approach has led to many interesting results: 
For example, scaling laws associated with the Laurent expansion of the 
map\cite{BDetal}, higher precision multifractal analysis 
\cite{BDetal2}, and the role of randomness\cite{BDetal3}.

It is well known that a flat interface becomes unstable in a 
diffusion-dominated growth process\cite{MS,SB}. 
The instability is suppressed within a certain 
length scale by the surface tension effect which stabilizes the flat interface
\cite{Langer}. Therefore the incorporation of this effect is important for 
considering the validity of a model being compared with experimental 
results and the relation between the randomness 
of a growth rule and regular growth patterns. 
There have been some attempts to incorporate the effect into the original 
form of the DLA, by introducing 
a sticking probability dependent on the curvature of the growth point
\cite{Vicsek,MFV} and more ingeniously, 
by reconstructing the interface by random walks between two points on 
the surface\cite{Kadanoff}.    

In the present article we present how to incorporate the surface tension 
effect into an iterative conformal mapping approach. 
We follow the idea of a probability dependent on the local curvature 
at the growth point\cite{Vicsek,MFV}. 
The curvature is calculated by utilizing the properties of the conformal map. 

The rest of the article is organized as follows: In Sec. 2, we describe 
the model. We briefly review the iterative conformal mapping approach. 
And then we introduce a probability with curvature dependence 
and present how to calculate 
the curvature. The patterns formed by this model are shown and qualitatively 
discussed in Sec. 3. A scaling analysis is given in Sec. 4. Sec. 5 
is dedicated to the summary.

\section{MODEL}
\subsection{Iterative Conformal Mapping Model of DLA}
Let $\Phi^{(n)}(z)$ be the conformal map which maps the external region 
of the unit circle $|z|=1$ in the complex plane $\mathbb C$ onto the 
external region of the two-dimensional $n-$particle cluster in the physical 
space. The image of the unit circle under $\Phi^{(n)}(z)$ is the interface of 
the cluster. 

First let $\phi_{\theta,\lambda}$ be the elementary map which makes a 
semicircular "bump" of size $\sqrt{\lambda}$ at around $\theta$ on the unit 
circle\cite{HL}:
\begin{eqnarray}
\phi_{\lambda,\theta}&=&e^{i\theta}\phi_{\lambda,0}(e^{-i\theta}z),
\\
\phi_{\lambda,0}(z)
&=&z^{1/2}\left\{(1+\lambda)\frac{1+z}{2z}
\left[1+z+\left(z^2+1-2z\frac{1+\lambda}{1-\lambda}\right)^{1/2}
\right]-1\right\}^{1/2}.
\end{eqnarray}
Then $\Phi^{(n)}(z)$ is constructed by iteration of the elementary maps:
\begin{equation}
\Phi^{(n)}(z) = \phi_{\theta_1,\lambda_1} \circ \phi_{\theta_2,\lambda_2}
\circ \cdots \circ\phi_{\theta_n,\lambda_n}(z).  
\end{equation}
Note the order of the composition.

The angle at which the $n$-th bump is generated, $\theta_n$, is an 
independent stochastic variable obeying the uniform distribution on 
$[0,2\pi]$. The bump size parameter $\lambda_n$ is chosen as
\begin{equation}
\lambda_n=\frac{\lambda_0}{|\Phi^{\prime}_{n-1}(e^{i\theta_n})|^2},
\label{bumpsize}
\end{equation}
in order to fix the bump size in the physical plane, to the first order, 
where $\lambda_0$ is a typical bump area and is specified in advance.

\begin{figure}
\begin{center}
\includegraphics[height=12.0cm]{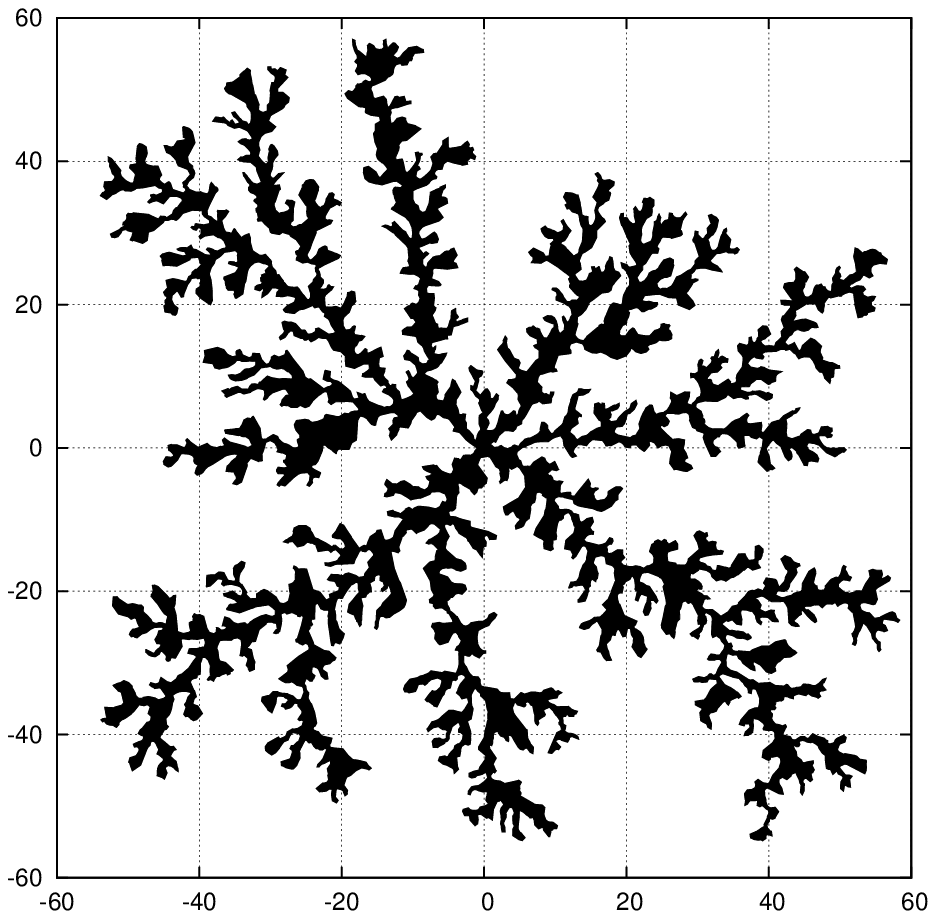}
%4.0cm for twocolumn format
\caption{\label{dla}}
\end{center}
\end{figure}

In Fig.\ref{dla} a typical cluster of 10000 bumps and $\lambda_0=0.1$  
constructed by the above approach is shown.  

\subsection{Surface tension effect}
In order to incorporate the surface tension effect into the conformal 
mapping model, we introduce the growth probability $P(z)$ 
that the next (($n+1$)-th) bump is generated at around   
$z=\Phi^{(n)}(e^{i\theta})$. 
The probability  $P(z)$ depends on the curvature at $z$ and is written as
\begin{equation}
P(z) = -A\kappa(z)+B,
\label{prob}
\end{equation}
where $A$ and $B$ are positive constants and $\kappa(z)$ is the curvature 
at $z$. Since $P(z)$ is defined as a probability, if $P(z)>1$ we set 
$P(z)=1$. Also, if $P(z)<\epsilon $ we set $P(z)=\epsilon$ where 
$\epsilon$ is a small positive constant, in order to keep the process going 
on and save the calculation time. It is necessary for $\epsilon$ to be 
smaller for larger $A$ in order to suppress the effect of randomness which 
allows more growth at a point with larger curvature. 

The introduction of the "sticking" probability, Eq.(\ref{prob}), that a 
random walker released far from the cluster sticks to a given point, to 
incorporate the surface tension effect, was presented in  
on-lattice\cite{Vicsek} and off-lattice\cite{MFV} simulations for 
diffusion-limited growth process.
Subsequently, it was shown\cite{Kadanoff} that
the evaluation of the surface tension effect could be better 
by reconstructing the interface by considering a random walk between two 
points on the interface. 
This method was used in simulations of 
viscous fingering on a lattice\cite{Liang,DBetal}. 
It was pointed out\cite{Sarkar} that there is a 
correspondence between  these two methods when we consider 
the short-time behavior of the interface not so far from flatness. 

In order to obtain the curvature at a point, 
we utilize the fact that the conformal map $\Phi^{(n)}$ 
has branch points associated with the generations of bumps\cite{BDP}. 
Each branch point has a preimage $\exp[i \beta^{j(n)}_{k,n}]$ 
on the unit circle. The argument of the preimage is 
characterized by three integer indices, $k$, $n$, and $j(n)$. The subscript 
$k$ is the step when the branch point was generated ({\it i.e.} the $k$-th 
bump was generated). The subscript $n$ is the step at which the analysis is 
being done. The superscript $j(n)$ is the order of the branch point along 
the arc length, emphasizing that it is a function of the step $n$. Consider 
the list of the arguments of the preimages at the $n$-th step, 
$\{ \beta^{j(n)}_{k,n}  \}$. For each step, the preimages move on the unit 
circle in the mathematical space keeping their physical positions fixed. 
When bumps overlap, some 
branch points are covered and their arguments are omitted from the list. 
Assume that the list at $(n-1)$-th step $\{ \beta^{j(n-1)}_{k,n-1}  \}$ is 
available. At the $n$-th step, the elementary map associated with the 
$n$-th bump $\phi_{\theta_n,\lambda_n}$ has two branch points at
\begin{equation}
\exp [i \alpha_n^{\pm}] = \phi_{\theta_n,\lambda_n}(\exp[i \beta^{\pm}_{n,n}]),
\end{equation} 
where
\begin{equation}
\beta^{\pm}_{n,n} = \theta_n \pm \tan^{-1}
\left[
\frac{2\sqrt{\lambda_n}}{1-\lambda_n}
\right],
\end{equation}
in order to keep the asymptotic form $\phi_{\theta_n,\lambda_n}(z) \sim z$ 
for $z \rightarrow \infty$. On the unit circle, the domain 
$[\alpha_n^-,\alpha_n^+]$ is covered by the bump. If an element of the list 
is in this domain, it does not join the list at the next step. Therefore 
the list at the $n$-th step $\{ \beta^{j(n)}_{k,n}  \}$ consists of the new 
elements $\beta^{\pm}_{n,n}$ and the elements of the list at the previous step, 
which are not omitted by the new bump and are updated as
\begin{equation} 
\exp[i \beta^{j'(n)}_{k,n}] = 
\phi_{\theta_n,\lambda_n}^{-1}(\exp[i \beta^{j(n-1)}_{k,n-1}]),
\end{equation}
where $\phi_{\theta_n,\lambda_n}^{-1}$ denotes the inverse function of
$\phi_{\theta_n,\lambda_n}$. 
Note that the order index $j'(n)$ is not simply related to 
$j(n-1)$. Each branch point $z_j$ at the $n$-th step is given as
\begin{eqnarray}
z_j &=& \Phi^{(n)}(\exp[i\beta^{j(n)}_{k,n}])
\nonumber
\\
&=& \Phi^{(k)} (\exp[i\beta^{\tilde{j}(n)}_{k,k}]),
\label{mapping}
\end{eqnarray} 
where $\exp[i\beta^{j(n)}_{k,n}]$ and $\exp[i\beta^{\tilde{j}(n)}_{k,k}]$ are 
related as
\begin{equation}
\exp[i\beta^{j(n)}_{k,n}] 
= \phi_{\theta_n,\lambda_n}^{-1} \circ \cdots \circ\phi_{\theta_{k+1},\lambda_{k+1}}^{-1}
(\exp[i\beta^{\tilde{j}(n)}_{k,k}]).
\end{equation}
Numerically the second-line representation of Eq.(\ref{mapping}) 
gives a much more precise result
than the first. We can draw the outline of the cluster by 
connecting the images of the branch points.

\begin{figure}{10}
\begin{center}
\includegraphics[height=12.0cm]{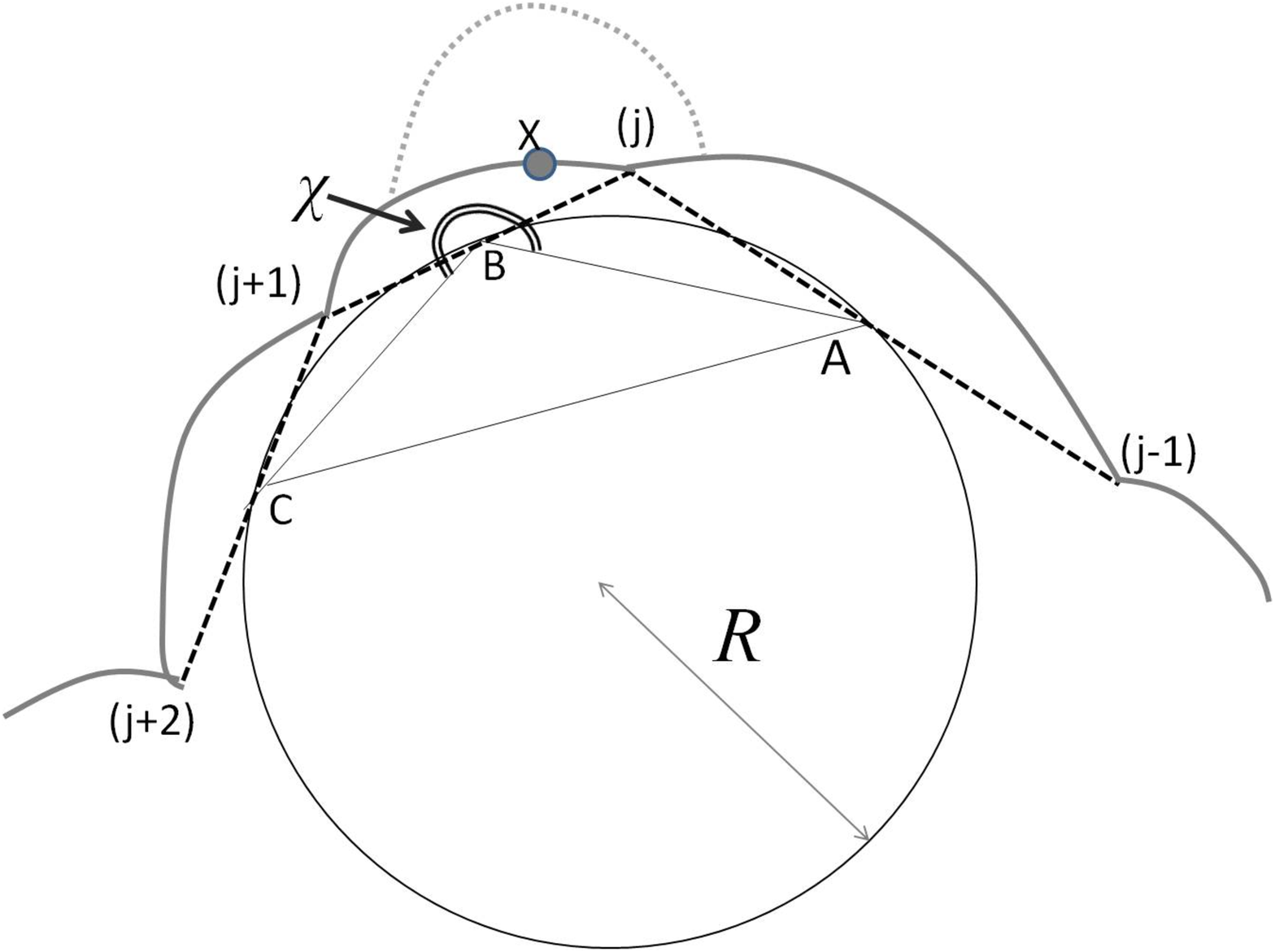}
%4.0cm for twocolumn format
\caption{\label{curvature}}
\end{center}
\end{figure}

Assume that $X=\Phi^{(n)}(e^{i\theta})$ is a candidate position of the
$n+1$-th bump which is between the $j$- and $(j+1)$-th branch points.
Let $A$, $B$, and $C$ be the midpoints between the $(j-1)$- and $j$-th, 
$j$- and $(j+1)$-th, and $(j+1)$- and $(j+2)$-th branch points, 
respectively. Then the curvature radius $R$ is given as the radius of the 
circumcircle of $\Delta ABC$ (see Fig.\ref{curvature}). 
Let $\chi=\angle ABC$ 
be the angle pointing to the external region. If $\chi > \pi$, the curvature 
is positive and if $\chi < \pi$ it is negative. The validity of this 
definition of the curvature will be examined in the next section.

For this model, it is known that sometimes a large flat bump is generated 
which seals a "fjord" of the cluster\cite{BDetal}. In order to prevent a 
flat bump from being generated, we restrict the number of branch points, 
$N_c$, covered by a bump. However, too small $N_c$ makes it impossible for a 
bump to be generated at a point of negative curvature, since branch points are 
concentrated around such point. Hereinafter, we set $N_c=6$.

\section{RESULTS}
Figs. \ref{plots}(a)-(d) show typical patterns for several values 
of $A$. Values $B=0.5$, $\lambda_0=0.1$ and $N=10000$ are common to all the 
calculations. 
The dependence on $B$ is very weak. The appropriate value of $\epsilon$ 
depends on $A$, since it should be chosen in order not to affect the results. 

The parameter $A$ corresponds to the strength of the surface tension effect. 
Therefore, with increasing $A$, the structure of the cluster becomes more 
coarse and compact and the cluster has fatter branches. 

We can also observe the above-described crossover of the pattern 
by changing the size of the cluster while keeping $A$ fixed, 
since the surface tension is effective within a 
certain finite length scale. This is demonstrated in Fig.\ref{growth}, 
where $A=2.0$, $B=0.5$, $\epsilon=0.0001$, and interfaces are plotted every 
750 steps. We can observe that an initially two-dimensional compact 
cluster becomes fractal through tip-splitting. 

In order for our definition of curvature to be valid, the curvature radius 
$R$ should be sufficiently larger than the distance between branch points. 
Fig. 5 shows the distribution of the curvature of the point at which 
a bump is generated for several values of $A$ until $N=10000$. 
A dependence of the distribution on step $n$ is not observed. It is easily 
confirmed that the greater $A$ becomes, the more suppressed the growth at 
a point with large curvature is. The characteristic distance between 
branch points is evaluated by the typical size of bump, 
$\sqrt{\lambda_0} \sim 0.3$. Thus it is necessary that a growth at a point 
with curvature larger than $\sim 3$ should not be allowed. 
From Fig.\ref{crvdist}, 
it is concluded that for $A \gtrsim 0.25$, our definition of curvature is 
valid.

\section{SCALING ANALYSIS}
In this model, some quantities can be evaluated from the conformal map. 
The area of the cluster $S_N$ is evaluated by the number of the bumps 
(steps) $N$, since the size of the bump is kept fixed to the first order 
(see Eq.(\ref{bumpsize})). The perimeter length $L_N$ of the cluster 
is evaluated by the number of the branch points $N_b$. From the Laurent 
expansion of the map $\Phi^{(n)}(z)$:
\begin{equation}
\Phi^{(n)}(z) 
= F_N^{(1)}z +F_N^{(0)}+\sum_{j=1}^{\infty} F_N^{(-j)}z^{-j},  
\end{equation}
the characteristic radius of the cluster of $N$ bumps $R_N$ is given by the 
first order coefficient, $ F_N^{(1)} $, which is called the Laplace radius 
or conformal radius and written in terms of the bump size parameter 
$\{ \lambda_j\}$ as
\begin{equation}
R_N \sim F_N^{(1)} \sim \prod^N_{j=1} (1+\lambda_j)^{1/2}.
\label{radiusclst}
\end{equation}
Note that $F_N^{(1)}$ can be taken to be positive without loss of generality.

We expect that the scaling relations shown below hold for these 
quantities:
\begin{eqnarray}
S_N \sim R_N^{D_S},
\\
L_N \sim R_N^{D_L},
\end{eqnarray}
where $D_s$ and $D_L$ are scaling exponents.

Figs.\ref{scaling} (a) and (b) show the plot of $R_N$ vs. $S_N$ 
and $R_N$ vs. $L_N$, respectively for $A=1.0$, averaged over 20 samples.
For $R_N \gtrsim R_{cN} \sim 5$, the growth is fractal, with the exponents 
$D_S =D_L \approx 1.7$. In this regime $D_S$ and 
$D_L$ are the fractal dimension of the area and  perimeter length, 
respectively. Our result $D_S=D_L \approx 1.7$ agrees with the result for 
the original DLA\cite{AMS}. For $R_N \lesssim R_{cN}$, the growth is 
compact, $D_S=2$ and $D_L=1$ (see Fig.\ref{growth}). If a larger $A$ is used 
and an average over more samples are taken, $R_{cN}$ becomes larger and the 
compact growth is expected to be observed more clearly. 

\section{SUMMARY}
We presented a method for incorporating the surface tension effect into an 
iterative conformal mapping model of the DLA. In our method, 
a generation of a bump depends 
on the curvature of the candidate point where this curvature is obtained by 
utilizing the branch points of a conformal map. Scaling analysis of the  
cluster grown under the proposed model is easily conducted 
using the conformal map technique. The grown pattern exhibited 
a crossover at a certain length scale from compact growth to  
fractal growth, due to the surface tension effect. In the fractal regime, the 
fractal dimensions of the area and the perimeter length coincide and their 
values agreed with those of previous results.

\section*{ACKNOWLEDGMENT}
This research was supported by the Japan Ministry of Education, Culture, 
Sports, Science and Technology, Grant-in-Aid for Scientific Research, 
No. 21540392.

\newpage
Figure captions:

Fig.1: Typical DLA cluster constructed by the iterated conformal 
mapping procedure, $N=10000$ and $\lambda_0=0.1$.
 
Fig.2: The approximate curvature at $X$, a candidate of the next growth 
point. The thick grey curve is the interface, which is the image of the unit 
circle under $\Phi^{(n)}$. The branch points of $\Phi^{(n)}$ are labelled along 
the arc length, $\cdots$, $(j-1)$,...,$(j+2)$, $\cdots$.  $A$, $B$ and $C$ are 
the midpoints between the  $(j-1)-$ and $j-$th, $j$ and $(j+1)-$ and 
$(j+1)-$ and $(j+2)-$th branch points, respectively. The radius of the 
circumcircle of $\Delta ABC$, denoted by $R$, is the approximate curvature 
radius at $X$. In this figure, the angle $\chi$ is the exterior angle of 
$\angle ABC$ and so the curvature at $X$ is positive.

Fig.3: Typical clusters for (a) $A=0.5$, $\epsilon=0.01$; (b) $A=1.0$, 
$\epsilon=0.001$; (c) $A=2.0$, $\epsilon=0.0001$; and (d) $A=4.0$, 
$\epsilon=0.0001$. $N=10000$ and $\lambda_0=0.1$ are common values.

Fig.4: Interfaces of the cluster of Fig.3(c). The contours are 
plotted every 750 steps. 

\begin{figure}[htbp]
\begin{minipage}{0.5\hsize}
\begin{center}
\includegraphics[width=8.0cm]{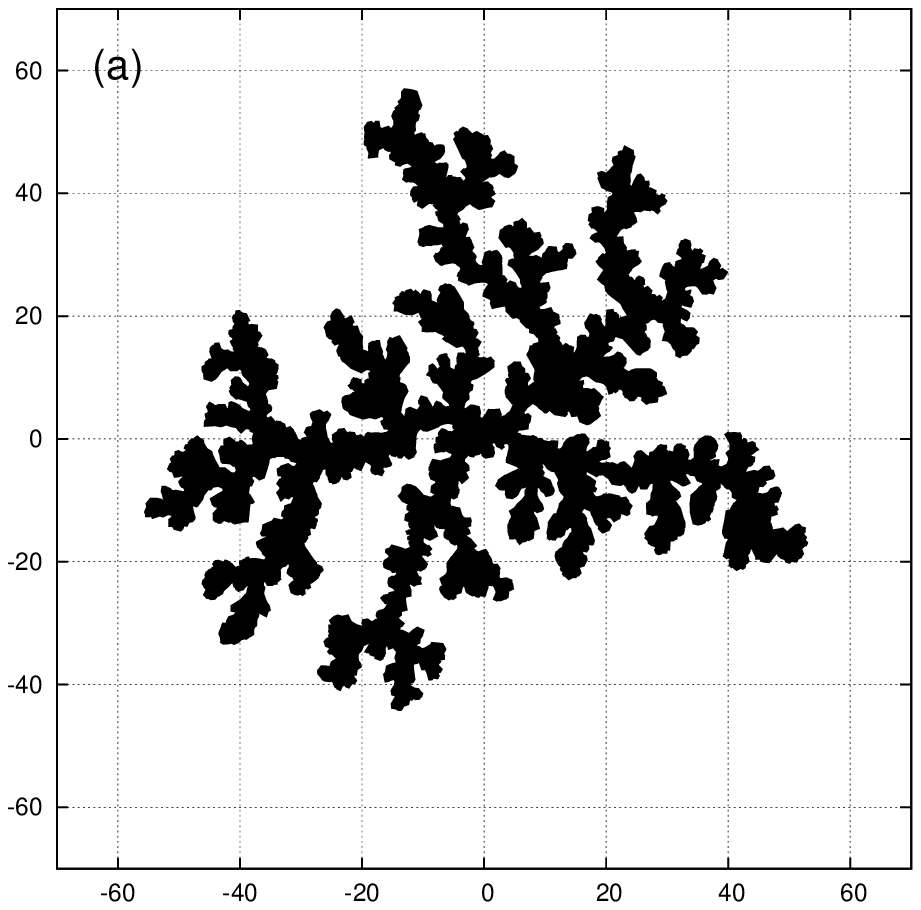}
\end{center}
\end{minipage}
\begin{minipage}{0.5\hsize}
\begin{center}
\includegraphics[width=8.0cm]{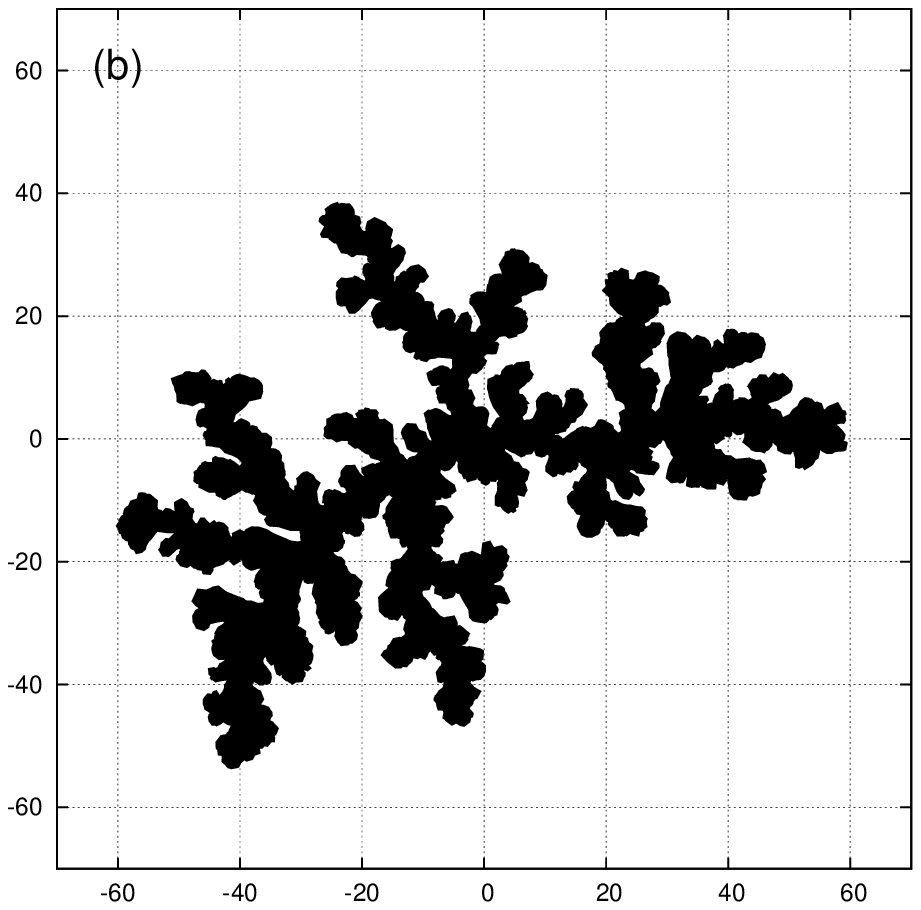}
\end{center}
\end{minipage}
\begin{minipage}{0.5\hsize}
\begin{center}
\includegraphics[width=8.0cm]{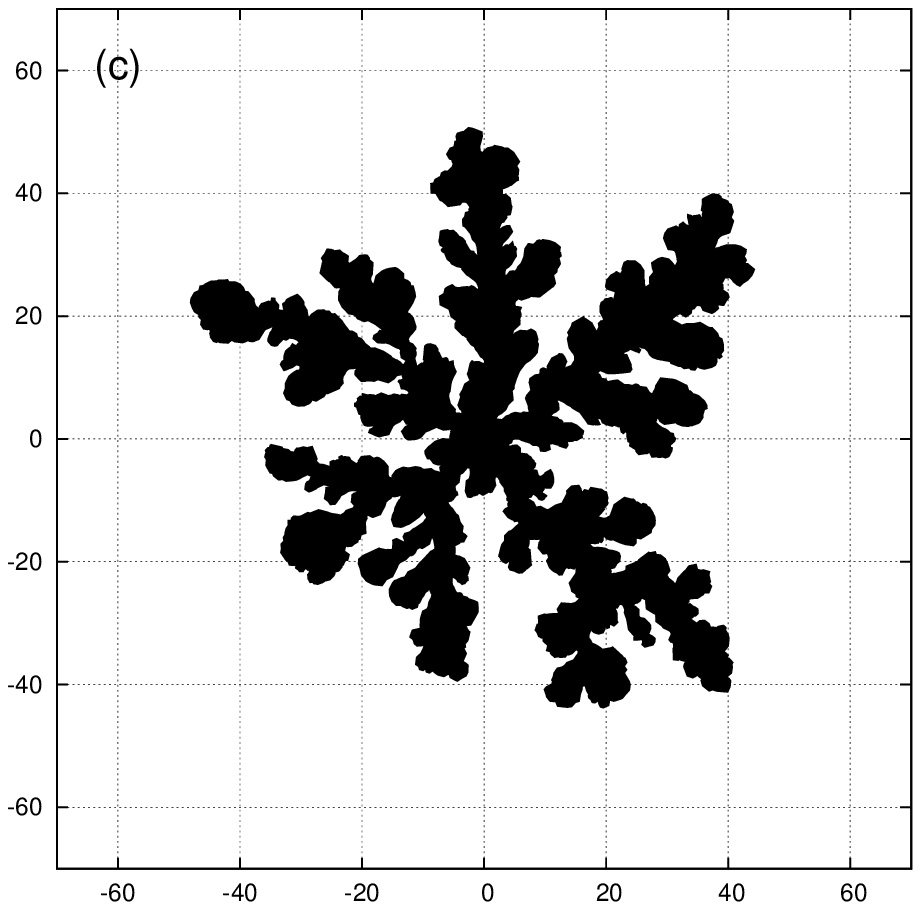}
\end{center}
\end{minipage}
\begin{minipage}{0.5\hsize}
\begin{center}
\includegraphics[width=8.0cm]{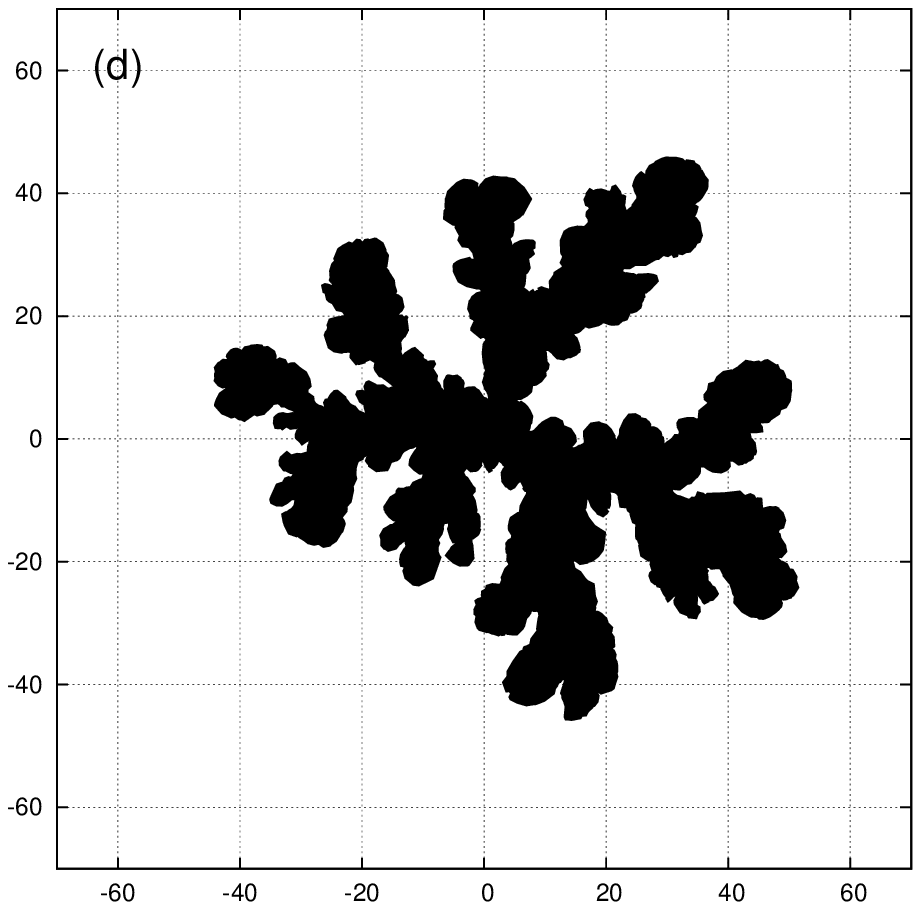}
\end{center}
\end{minipage}
\caption{\label{plots}}
\end{figure}

\begin{figure}
\begin{center}
\includegraphics[height=12.0cm]{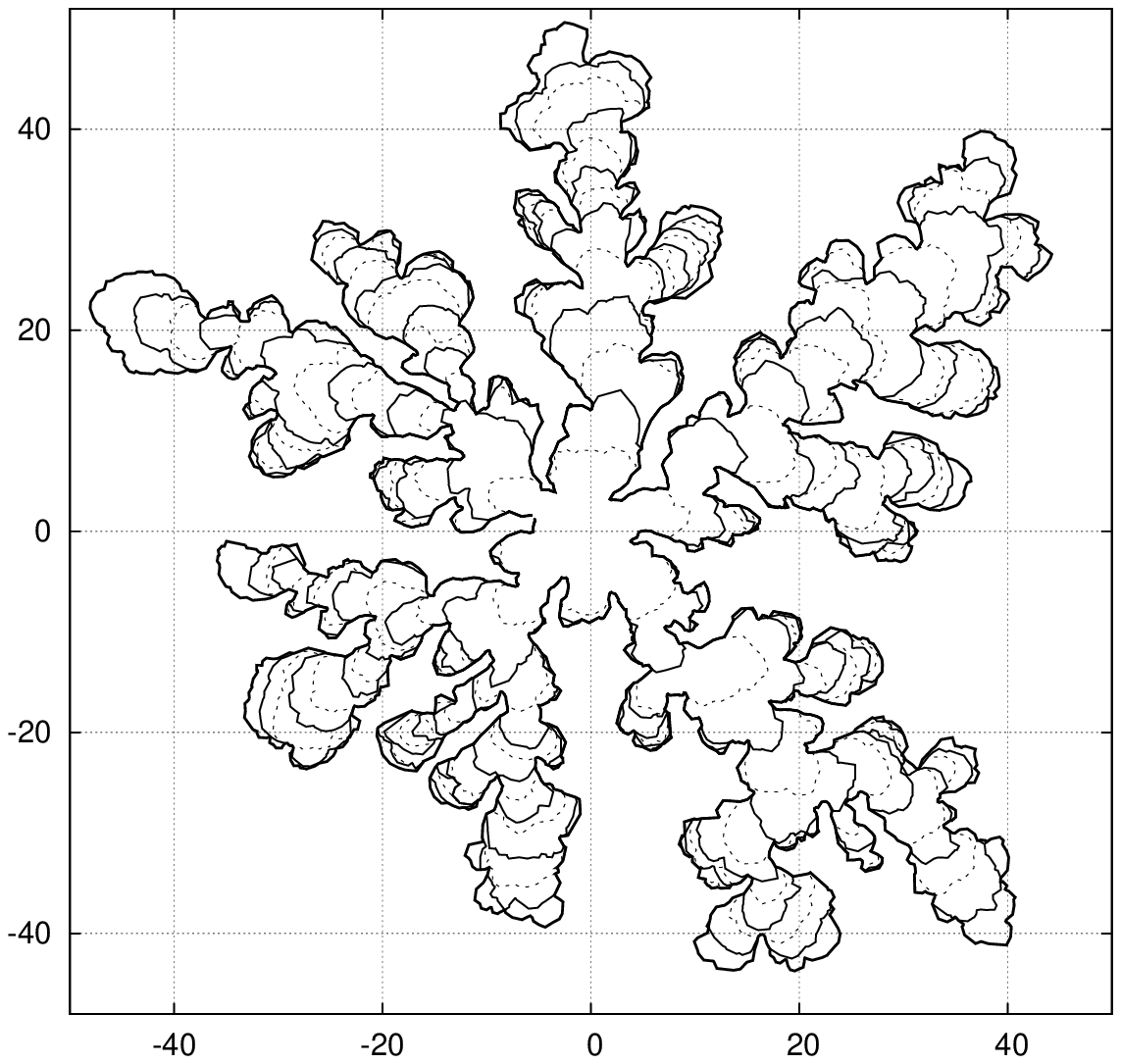}
%4.0cm for twocolumn format
\caption{\label{growth}}
\end{center}
\end{figure}

\begin{figure}
\begin{center}
\includegraphics[height=12.0cm]{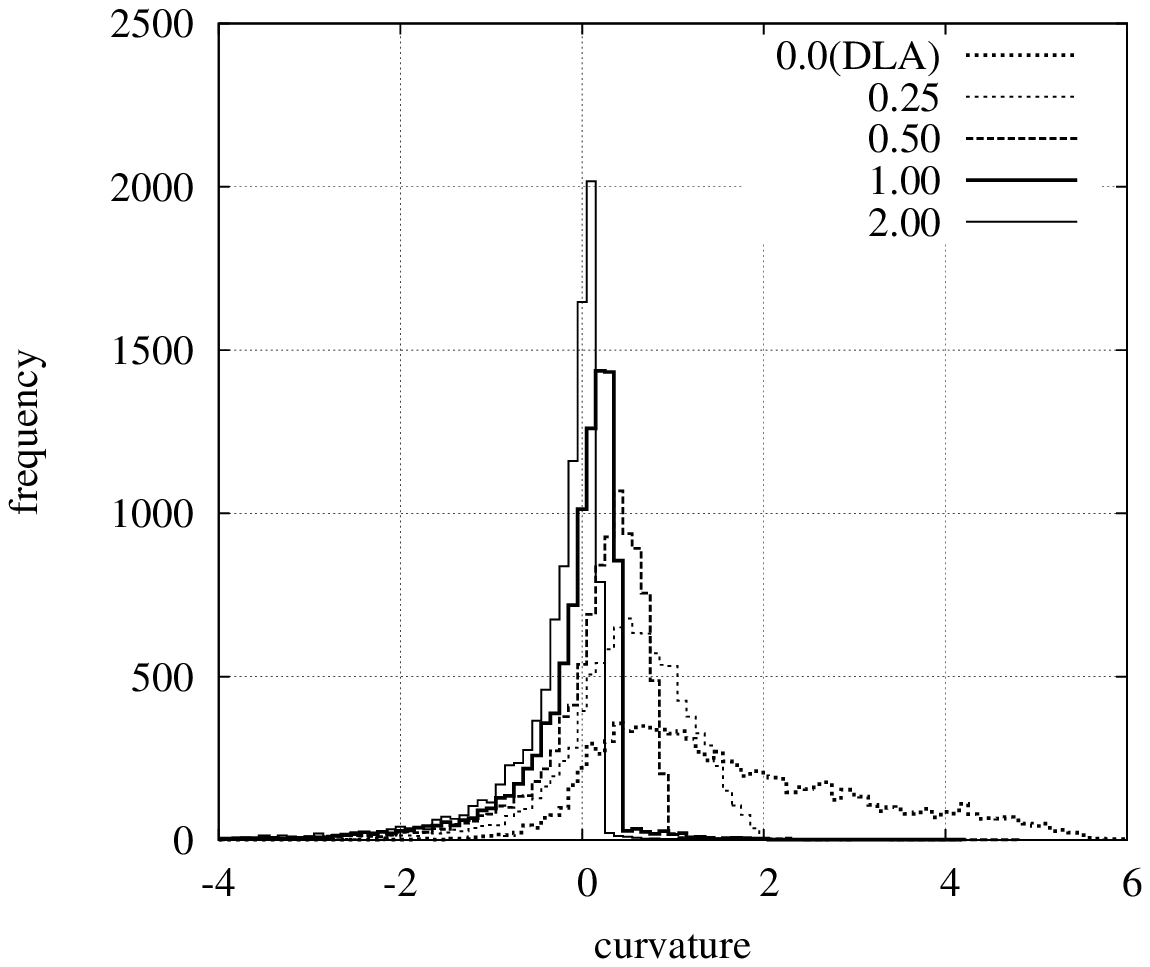}
%4.0cm for twocolumn format
\caption{\label{crvdist}}
\end{center}
\end{figure}

\begin{figure}
\centering
\begin{minipage}{0.75\hsize}
%\begin{center}
\includegraphics[height=10.0cm]{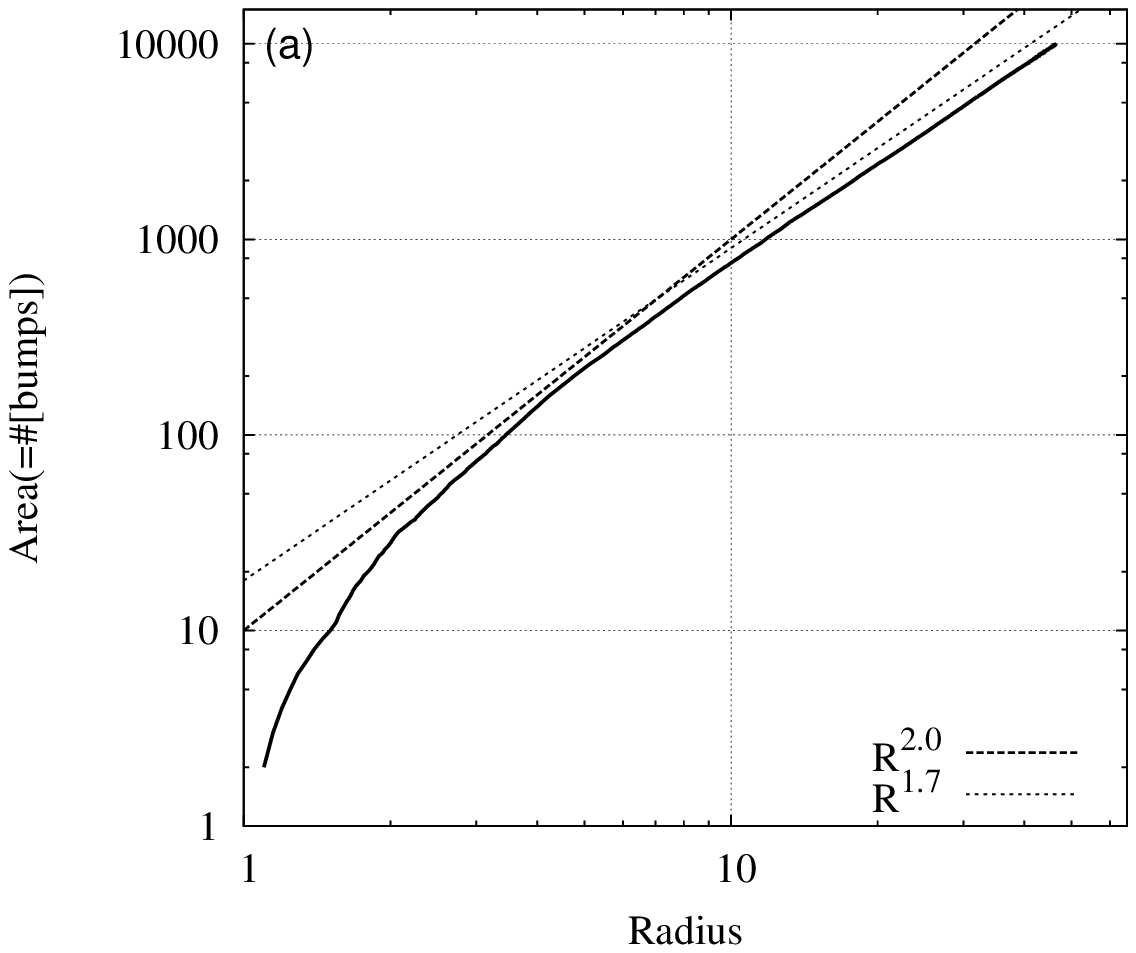}
%\end{center}
\end{minipage}
\begin{minipage}{0.75\hsize}
%\begin{center}
\includegraphics[height=10.0cm]{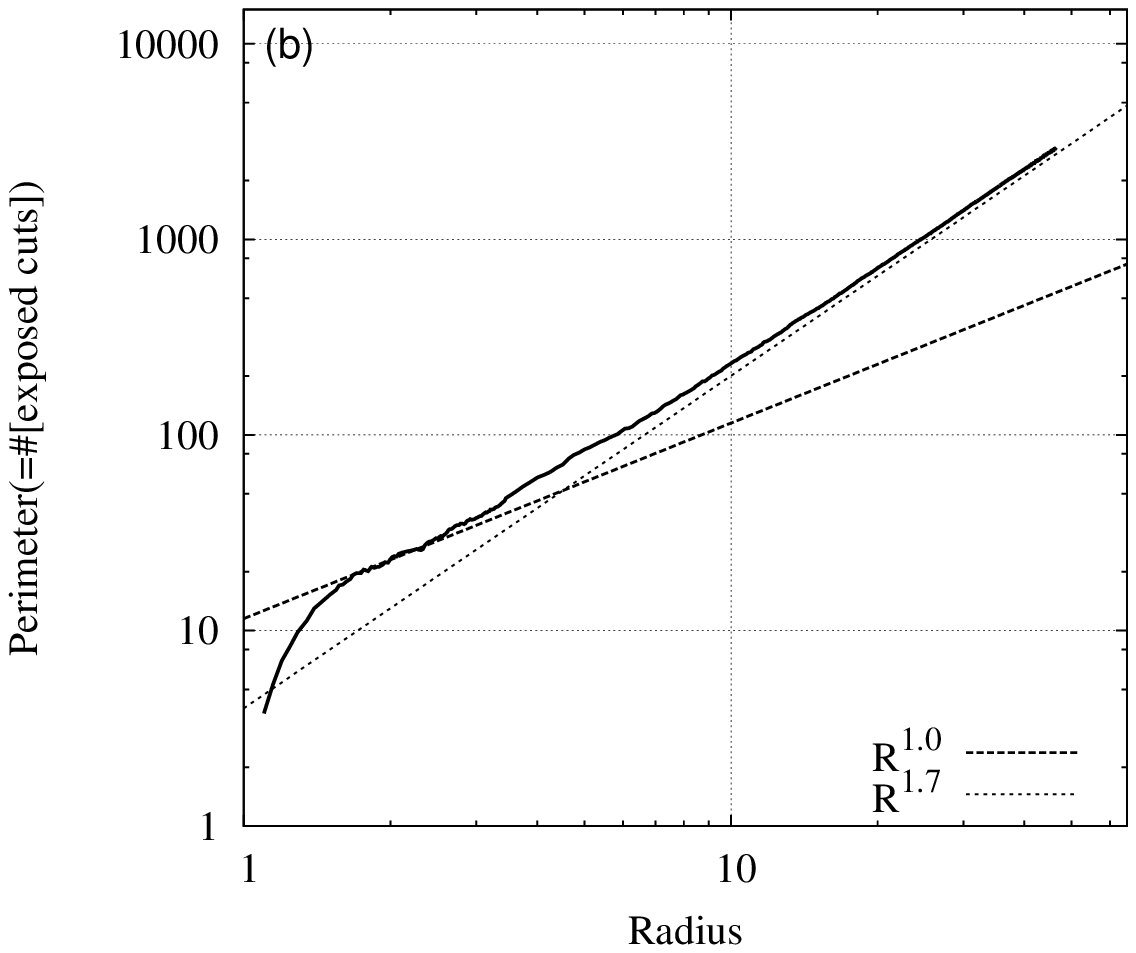}
%\end{center}
\end{minipage}
\caption{\label{scaling}}
\end{figure}

Fig.5: Distribution of the curvature of growth points for 
several values of the strength of surface tension effect $A$. 
The values of $\epsilon$ are the same as those for
Fig.\ref{plots}, with the addition of $\epsilon=0.01$ for $A=0.25$.

Fig.6: (a)The log-log plot of the cluster area $S_N$ versus radius $R_N$. 
(b)The log-log plot of the perimeter length $L_N$ versus $R_N$ for $A=1.0$, 
$\epsilon=0.001$. In each plot the broken and dotted lines are guides for 
the eyes.
\newpage

\begin{thebibliography}{99}
\bibitem{WS}
T.A.Witten and L.M.Sander, Phys.Rev.Lett.{\bf 47}(1981)1400-1403;
Phys.Rev.B {\bf 27}(1981)5686-5697.

\bibitem{Mat}
M.Matsushita {\it et al}., Phys.Rev.Lett. {\bf 53}(1984)286-289.

\bibitem{HS}
J.H.S.Hele-Shaw, Nature {\bf 58}(1898)34-36.

\bibitem{ST}
P.G.Saffman and G.I.Taylor, Proc.Roy.Soc.Lond.A {\bf 245}(1958)312-329.

\bibitem{NDS}
G.Daccord, J.Nittman and H.E.Stanley, 
Phys.Rev.Lett. {\bf 56}(1986)336-339.

\bibitem{HOM}
H.Honjo, S.Ohta and M.Matsushita, J.Phys.Soc.Jpn. {\bf 55}(1986)
2487-2490.

\bibitem{FM}
H.Fujisaka and M.Matsushita, J.Phys.Soc.Jpn. {\bf 58}(1989)
3875-3878.

\bibitem{HL}
M.B.Hastings and L.S.Levitov, Physica D {\bf 116}(1998)244-252;
M.B.Hastings, Phys.Rev.E {\bf 55}(1997)135-152.

\bibitem{BDetal}
B.Davidovitch {\it et al}., Phys.Rev.E {\bf 59}(1999)1368-1378.

\bibitem{BDetal2}
B.Davidovitch {\it et al}., Phys.Rev.Lett. {\bf 87}(2001)164101;
M.H.Jensen {\it et al}., Phys.Rev.E {\bf 65}(2002)046109.

\bibitem{BDetal3}
B.Davidovitch {\it et al}., Phys.Rev.E {\bf 62}(2000)1706-1715.

\bibitem{MS}
W.W.Mullins and R.F.Sekerka, J.Appl.Phys. {\bf 34}(1963)323-329; 
{\it ibid}. {\bf 35}(1964)444-452.

\bibitem{SB}
B.Shraiman and D.Bensimon, Phys.Rev.A {\bf 30}(1984)2840-2842.

\bibitem{Langer}
J.S.Langer, Rev.Mod.Phys. {\bf 52}(1980)1-28.

\bibitem{Vicsek}
T.Viscek, Phys.Rev.Lett. {\bf 53}(1984)2281-2284;
Phys.Rev.A {\bf 32}(1985)3084-3089.

\bibitem{MFV}
P.Meakin, F.Family and T.Vicsek, 
J.Colloid and Interface Sci. {\bf 117}(1987)394-399.

\bibitem{Kadanoff}
L.P.Kadanoff, J.Stat.Phys. {\bf 39}(1985)267-283.

\bibitem{Liang}
S.Liang, Phys.Rev.A {\bf 33}(1986)2663-2674.

\bibitem{DBetal}
D.Bensimon {\it et al}. Rev.Mod.Phys. {\bf 58}(1986)977-999.

\bibitem{Sarkar}
S.K.Sarkar, Phys.Rev.A {\bf 32}(1985)3114-3116.

\bibitem{BDP}
F.Barra, B.Davidovitch and I.Procaccia,
Phys.Rev.E {\bf 65}(2002)046144. 

\bibitem{AMS}
C.Amitrano, P.Meakin and H.E.Stanley,
Phys.Rev.A {\bf 40}(1989)1713-1716.

\end{thebibliography}
\end{document}